\documentclass[prb,twocolumn,showpacs,superscriptaddress,floatfix]{revtex4}

\usepackage{graphicx}
\usepackage{amsmath}
\usepackage{amssymb}
\usepackage{bm}

\begin{document}


\title{
Anisotropy on the Fermi Surface
of the Two-Dimensional Hubbard Model
}

\author{Y. Otsuka}
\email{otsuka@pothos.t.u-tokyo.ac.jp}
\affiliation{
Department of Applied Physics, University of Tokyo,
7-3-1 Hongo Bunkyo-ku,
Tokyo 113-8656, Japan
}

\author{Y. Morita}
\affiliation{
Department of Applied Physics, University of Tokyo,
7-3-1 Hongo Bunkyo-ku,
Tokyo 113-8656, Japan
}

\author{Y. Hatsugai}
\affiliation{
Department of Applied Physics, University of Tokyo,
7-3-1 Hongo Bunkyo-ku,
Tokyo 113-8656, Japan
}
\affiliation{
PRESTO,
Japan Science and Technology Corporation,
Kawaguchi-shi,Saitama, 332-0012, Japan
}

\date{\today}

\begin{abstract}
We investigate anisotropic charge fluctuations 
in the two-dimensional Hubbard model at half filling.
By the quantum Monte Carlo method, 
we calculate a momentum-resolved charge compressibility 
$\kappa (\bm{k}) = {d \langle n(\bm{k}) \rangle }/{d \mu}$, 
which shows effects of an infinitesimal doping.
At the temperature $T \sim {t^2}/{U}$,
$\kappa (\bm{k})$ shows peak structure 
at  the $(\pm \pi/2,\pm \pi/2)$ points
along the $|k_x| + |k_y| = \pi$ line.
A similar peak structure is reproduced 
in the mean-filed calculation 
for the $d$-wave pairing state or the staggered flux state.
\end{abstract}


\maketitle


Effects of electron-electron interaction have 
attracted lots of interest.
One of the most significant phenomena 
in the strongly correlated electron system
is the Mott transition,
which is a quantum phase transition
driven by the interaction.
This Mott insulator has a finite charge gap
and the anitiferromagnetism often accompanies.
Recently, various anomalous properties have been found
near the Mott transition.
Especially, there has been a proposal that
the interaction brings about anisotropy
in the low energy excitations.
For example,
the spectral weight of 
the two-dimensional Hubbard model
has been investigated~\cite{bulut94-b1,bulut94-b2,hanke,ift-rmp}.
In addition,
the deformation of the Fermi surface due to the interaction
occurs in the $t-J$ model~\cite{putika98,ogata00}.
The singular momentum dependence is also observed experimentally.
Angle resolved photoemission spectroscopy (ARPES) measurements
suggest anisotropic properties 
in the low energy excitations~\cite{ronning98}.
These anomalous properties are 
widely observed in the strongly correlated electron system
and they could be an evidence of the non-Fermi liquid behavior.
It is thus an important and appealing topic to study
the low energy excitations with momentum resolution.

The anisotropy in the low energy excitations is natural, 
if we assume 
the $d$-wave pairing state or the flux state~\cite{affleck88-prb37}.
They both have a singular energy dispersion that has gap nodes
along the diagonal directions $|k_{x}|=|k_{y}|$, while a gap opens
around the $(\pm \pi,0)$ and $(0, \pm \pi)$ points.
Thus, one expects anisotropic charge excitations near the Fermi surface.
In general, however, these states compete with other instabilities.
Especially, at a rational filling,
the system often belongs to 
the antiferromagnetic Mott insulator,
where the charge degree of freedom is frozen.
In that case, the N\'{e}el state well describes the ground state.
Nevertheless, 
the $d$-wave pairing state or the flux state can give
a sound basis for the interpretation of 
some singular phenomena
near the Mott insulator at a finite temperature.

 In this letter, 
we investigate the charge fluctuation 
in the half-filled Hubbard model
on a two-dimensional square lattice.
Using the quantum Monte Carlo method,
we calculate the momentum-resolved charge compressibility
\begin{equation}
\kappa(\bm{k}) = \frac{d \langle n(\bm{k}) \rangle}{d\mu} \nonumber ,
\label{eq:kappa}
\end{equation}
where 
$\langle n(\bm{k}) \rangle =\langle c_{\bm{k}}^{\dagger} c_{\bm{k}} \rangle$
denotes the momentum distribution function
and $\mu$ the chemical potential.
The integral of $\kappa(\bm{k})$ is equal to
the charge compressibility 
$ \kappa = \int d\bm{k} \kappa(\bm{k})$.
If system has a finite charge gap,
the charge compressibility 
decreases exponentially towards zero
as a temperature is lowered.
Thus, at half filling, 
we cannot calculate $\kappa(\bm{k})$ 
with sufficient numerical accuracy
at a very low temperature.
At a finite temperature,
an infinitesimal shift of $\mu$ causes
an infinitesimal doping.
This means 
$\kappa(\bm{k})$ reflects 
the distribution of infinitesimally doped carriers 
with momentum resolution
$
 \delta  n(\bm{k}) \approx \kappa(\bm{k}) \delta  \mu \nonumber .
$
Without the interaction,
$\kappa(\bm{k})$ is peaked on the Fermi surface
and its value is constant.
At half filling, this non-interacting Fermi surface
is a square in the Brillouin zone,
$|k_x| + |k_y| = \pi$. 
We define the Fermi surface(FS) as this square in this paper.

We calculate $\kappa(\bm{k})$ at half filling.
Here the charge degree of freedom is almost frozen
and the system is dominated only by the insulating 
fixed point (antiferromagnetic Mott insulator)
at the low temperature($T \ll t^{2}/U$).
On the other hand, at a sufficiently high temperature($T \gg U$),
the Coulomb interaction $U$ is irrelevant.
We focus on the intermediate temperature region ($T \sim {t^2}/{U}$),
expecting  an interaction between charge and spin degree of freedom
gives non-trivial feature on $\kappa (\bm{k})$
even at half-filling.

 The Hamiltonian of the two-dimensional Hubbard model
is given by
\begin{align}
 \mathcal{H}
   =&
   -t \sum_{ \langle i,j \rangle , \sigma}
   \left(
   c_{i \sigma }^{\dagger} c_{j \sigma} 
   +
   c_{j \sigma}^{\dagger}  c_{i \sigma}
   \right)\nonumber\\
   &+
   U \sum_{i}
   \left( n_{i \uparrow  } - 1/2 \right)
   \left( n_{i \downarrow} - 1/2 \right)
   -
   \mu \sum_{i, \sigma} n_{i \sigma} \nonumber,
  \label{eq:Hamiltonian_Hubbard}
\end{align}
where 
$\langle i,j \rangle$ denotes the nearest-neighbor links,
and $t$ the nearest-neighbor hopping amplitude.
The system is on a square lattice and we impose periodic boundary conditions.

 In order to obtain approximation-free results, 
we employ the finite temperature auxiliary field
quantum Monte Carlo (QMC) method~\cite{bss,hirsch85,white89}.
In this method, physical observables are evaluated
in the grand canonical ensemble.
This makes it possible to obtain
$\kappa(\bm{k})$ by direct sampling in the QMC simulations as
\begin{equation}
\kappa (\bm{k})
= \beta
\left( 
\langle n(\bm{k}) \sum_{\bm{k}} n(\bm{k}) \rangle
-
\langle n(\bm{k}) \rangle \langle  \sum_{\bm{k}} n(\bm{k}) \rangle
\right) \nonumber ,
\end{equation}
where $\beta$ denotes an inverse temperature.
If we calculate $\kappa(\bm{k})$ by
numerically differentiating 
the QMC data $\langle n(\bm{k}) \rangle$,
we need to perform simulations for the doped system.
This brings about the notorious sign problem,
which prevents us from getting reliable data. 
On the other hand,
the direct evaluation of $\kappa(\bm{k})$
only needs a simulation at half filling,
where the sign problem does not occur
due to the particle-hole symmetry.
The number of the electrons is not fixed at half filling
in the QMC ensembles,
even if we set the chemical potential $\mu=0$.
Thus,
the information on infinitesimally doped systems is 
statistically taken into account in $\kappa(\bm{k})$.

 The simulations were performed on a $16 \times 16$ square lattice.
The finite size effects on $\kappa(\bm{k})$ are not observed
with this lattice size at the temperatures we have studied.
The Trotter time slice size is set to be $\Delta\tau \simeq 0.1/t$.
We have checked that the systematic error due to 
the Trotter decomposition
does not change qualitative features.
For the interacting case, the strength of the interaction 
is set to be $U/t=4$, 
where the charge gap is $E_g \simeq 0.6t$~\cite{Furukawa}.
We have typically performed 500 Monte Carlo sweeps
in order to reach a thermal equilibrium 
followed by $\sim 10^4$ measurement sweeps.
The measurements are divided into 10 blocks 
and the statistical error is estimated
by the variance among the blocks.

\begin{figure}[htbp]
 \begin{center}
  \includegraphics[width=9cm,clip]{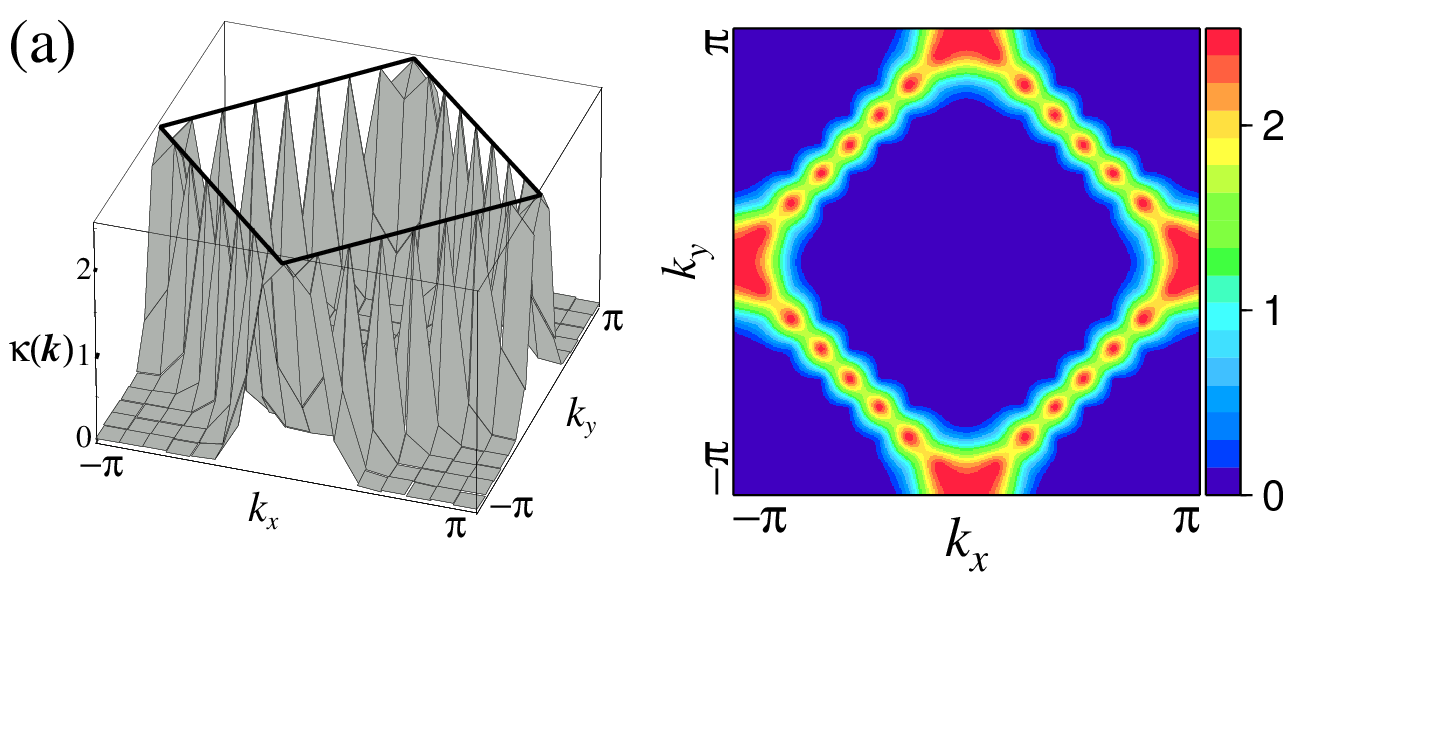}
  \includegraphics[width=9cm,clip]{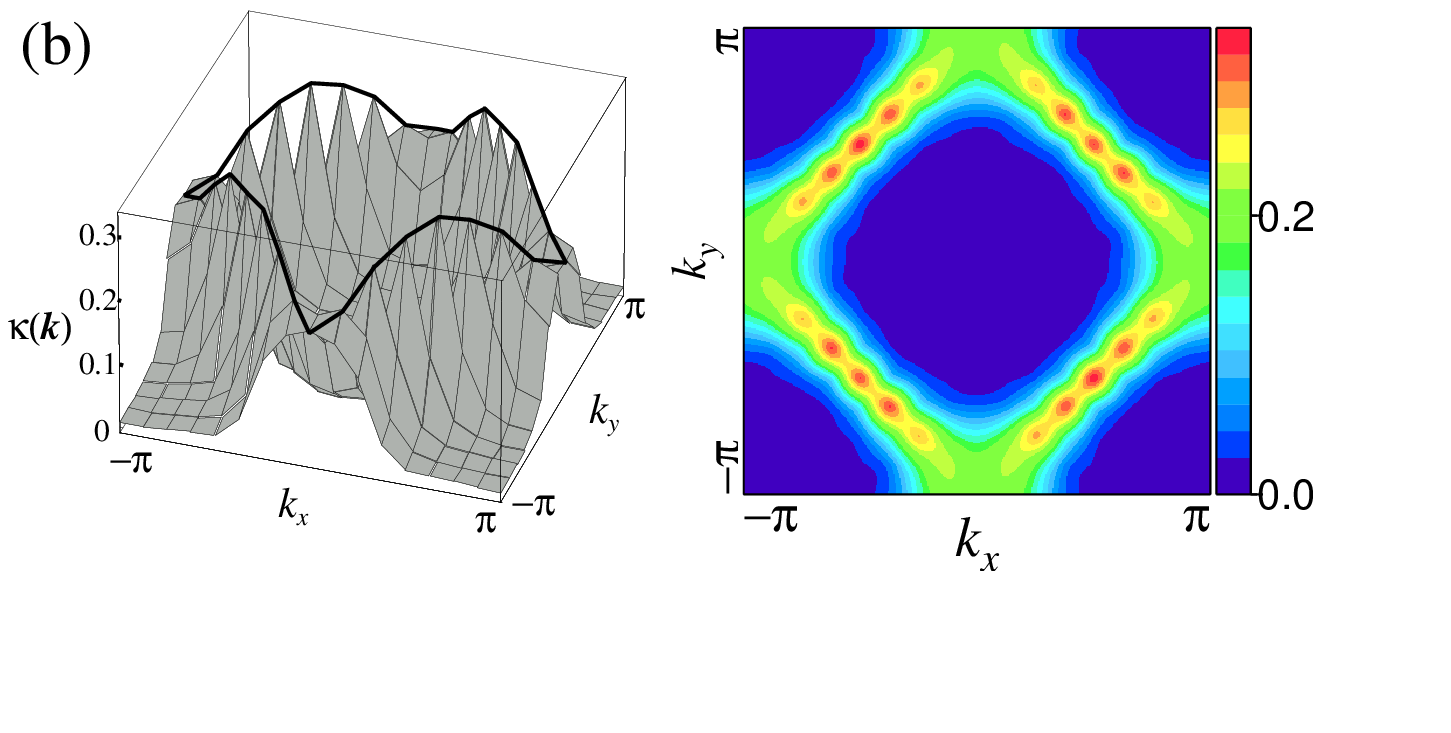}
  \caption{
    The momentum-resolved compressibility
  $\kappa (\protect\bm{k}) = {d \langle n(\protect\bm{k}) \rangle }/{d \mu}$
  obtained by the quantum Monte Carlo simulations
  at $T/t=0.2$ on a $16 \times 16$ lattice:
  (a) for $U/t=0$;
  (b) for $U/t=4$.
  The areas with large values of $\kappa (\protect\bm{k})$ are
  highlighted for the contour plots.
    Without the interaction($U/t=0$),
  $\kappa (\protect\bm{k})$ is constant 
  on the Fermi surface. 
    On the other hand, for $U/t=4$,
  $\kappa (\protect\bm{k})$ shows  peak structure
  at the $(\pm\pi/2,\pm\pi/2)$ points.}
  \label{fig:qmc01} 
 \end{center}
\end{figure}

 At first, let us discuss results at 
the temperature $T \sim {t^2}/{U}$. 
The results of $\kappa(\bm{k})$
for $U/t=0$ and $U/t=4$ are shown in Fig.~\ref{fig:qmc01}.
For the non-interacting case, 
the value of $\kappa(\bm{k})$ on the FS is constant.
On the other hand, for $U/t=4$,
peak structure emerges in $\kappa(\bm{k})$ 
at the ($\pm\pi/2$,$\pm\pi/2$) points.
It indicates that
the ($\pm\pi/2$,$\pm\pi/2$) points are more sensitive 
to the shift of the chemical potential
than ($\pm\pi$,0) or ($0,\pm\pi$).
In other words,
the system is more {\it compressible} 
at the $(\pm\pi /2, \pm\pi /2)$ points.
Here we note that
the interaction does not change
the line defined by $\langle n(\bm{k}) \rangle =0.5$,
which is identical with the FS in the non-interacting case.
In that sense, 
the shape of the Fermi surface itself
is not deformed at half filling~\cite{Halboth}.

\begin{figure}[htbp]
 \begin{center}
  \includegraphics[width=8.0cm,clip]{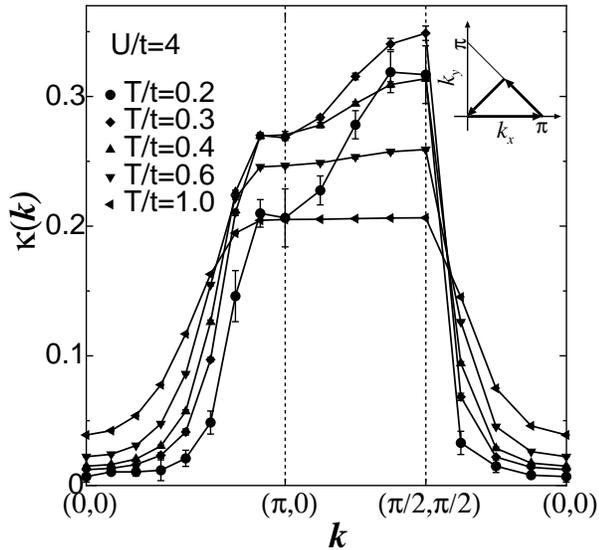}
 \caption{
    The momentum-resolved compressibility
  $\kappa (\protect\bm{k}) = {d \langle n(\protect\bm{k}) \rangle }/{d \mu}$
  obtained by the quantum Monte Carlo simulations
  on a $16 \times 16$ lattice
  for $U/t=4$
  at various temperatures.
    The peak structure at the $(\pm\pi/2,\pm\pi/2)$ points 
  vanishes at $T > U$.
  }
  \label{fig:qmc02} 
 \end{center}
\end{figure}

 Next, we discuss the temperature dependence
of $\kappa(\bm{k})$.
The results of $\kappa(\bm{k})$
for $U/t=4$ at several temperatures are provided in Fig.~\ref{fig:qmc02}.
The peak structure at the $(\pm\pi /2, \pm\pi /2)$ points is 
clearly observed at $T \sim {t^2}/{U} $.
It becomes ambiguous as the temperature increases 
and vanishes above $T \sim U$.
Here we note that
the antiferromagnetic correlation length
is smaller than the linear system size
at these temperatures ($T/t \ge 0.2$).
There are two known characteristic energy scales
in the Hubbard model.
One is the Coulomb interaction  $U$ and
the other is the effective superexchange interaction $J \sim {t^2}/{U}$.
Our results show the peak structure in $\kappa(\bm{k})$
emerges at the temperature $T \sim J$
and vanishes at $T > U$.

In the half-filled Hubbard model,
an antiferromagnetic long-range order 
appears at $T=0$.
At a finite temperature, 
the system does not have any long-range order.
Then there is no a priori reason 
to expect 
such an anisotropy in the charge compressibility.
Thus, let us compare the QMC results with various 
mean-field solutions at the temperature $T \sim J$.
We focus on three possible mean-field solutions;
the N\'{e}el state, the $d$-wave pairing state, and the staggered flux state.
As an effective model of the Hubbard model~\cite{Harris},
we use the following $t-J$ model to compare with our QMC results
\begin{align}
 \mathcal{H}
 =& 
 -t \sum_{{\langle}j,k{\rangle},{\sigma}}	
 ( c_{j{\sigma}}^{\dagger}c_{k{\sigma}} 
 + c_{k{\sigma}}^{\dagger}c_{j{\sigma}} ) \nonumber \\
 &+ J \sum_{{\langle}j,k{\rangle}} 
 \left(
 \bm{S}_{j} \cdot \bm{S}_{k}
 - \frac{1}{4} n_{j}n_{k}
 \right)
 - \mu \sum_{i,\sigma} n_{i \sigma} \nonumber,
\label{eq:t-J}
\end{align}
where
$
\bm{S}_{i}=
\frac{1}{2}
\sum_{{\sigma},{\sigma}^{\prime}}
c_{i{\sigma}}^{\dagger}
\bm{\sigma}_{{\sigma}{\sigma}^{\prime}}
c_{i{\sigma}^{\prime}}
$ 
and the double occupancy at the same site is prohibited.
We take order parameters as
\begin{align*}
\Delta_{jk} 
 &= 
 \frac{J}{2}
 \langle\left(
 c_{j{\uparrow}}c_{k{\downarrow}}
 {}- c_{j{\downarrow}}c_{k{\uparrow}}
 \right)\rangle
 & &\text{($d$-wave pairing),}\\
 \chi_{jk} 
 &= 
 \frac{J}{2}
 \langle\left(
 c_{j{\uparrow}}c_{k{\uparrow}}^{\dagger}
 + c_{j{\downarrow}}c_{k{\downarrow}}^{\dagger}
 \right)\rangle
& &\text{(staggered flux),}\\
m_{j}
&= J \langle S_{j}^{z} \rangle
& &\text{(N\'{e}el).}
\end{align*}
We set ${\Delta}_{jk}$ to have the $d$-wave symmetry 
and its amplitude is constant $\Delta$.
The link order ${\chi}_{jk}$ is chosen so that
the effective magnetic flux for each plaquette 
takes ${\phi}$ or $-{\phi}$ alternately.
The parameters satisfy 
${\Delta}/t < 1$ or $\phi < {\pi}$
where the low-lying excitations are described by 
anisotropic Dirac fermions\cite{morita01}.
The staggered magnetization $m_j$ is $(-1)^{j_x+j_y} m $.
In order to apply the mean-field ansatz,
we introduce auxiliary fields to decouple the superexchange
term and incorporate the constraints of no double occupancy. 
Next, we apply the saddle-point approximation and neglect the 
fluctuation around it by taking order parameters as an input parameter
(without self consistency condition).
In low dimensions, 
the fluctuations generally play a role of destructing 
the long-range order
and it is natural to expect it in our case.
Our focus is, however, 
not on such a long-distance behavior.

\begin{figure}[htbp]
 \begin{center}
  \includegraphics[width=5cm,clip]{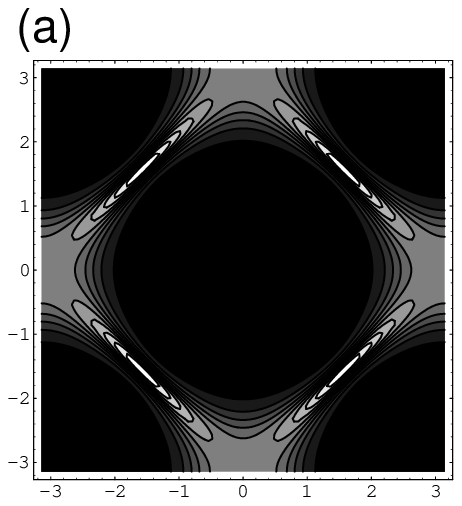}
  \includegraphics[width=5cm,clip]{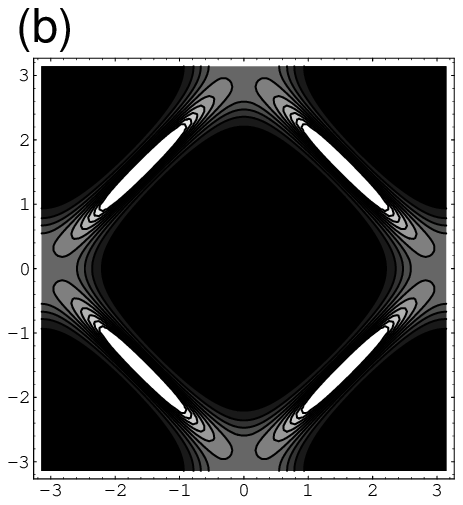}
  \includegraphics[width=5cm,clip]{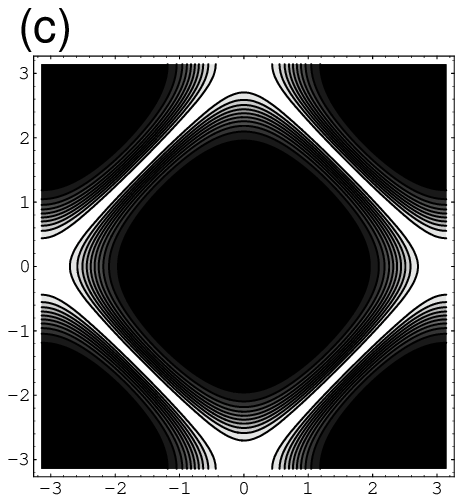}
  \caption{
  The mean-field calculations of 
  the momentum-resolved compressibility
  $\kappa (\protect\bm{k}) = {d \langle n(\protect\bm{k}) \rangle }/{d \mu}$
  at $T/t=0.2$:
  (a) for the $d$-wave pairing state($\Delta/t=0.2$);
  (b) for the staggered flux state ($\phi=\pi/6$);
  (c) for the N\'{e}el state($m/t=0.6$).
  The areas with large values of $\kappa (\protect\bm{k})$ are highlighted.
    The $\kappa (\protect\bm{k})$ shows peak structure 
  at the ($\pm \pi/2, \pm \pi/2$) points on the Fermi surface
  for the $d$-wave or staggered flux state,
  which is consistent with 
  the quantum Monte Carlo simulations.
    On the other hand, for the N\'{e}el state,
  $\kappa (\protect\bm{k})$ is constant
  on the Fermi surface.
  }
 \label{fig:mf}
 \end{center}
\end{figure}

We show $\kappa(\bm{k})$ at half filling 
for these mean-field states in Fig.~\ref{fig:mf}.
The peak structure at the ($\pm \pi/2$,$\pm \pi/2$) points
is observed in the $d$-wave pairing or the staggered flux state,
which is similar to the QMC results.
This implies 
the $d$-wave pairing state or the staggered flux state
competes with the N\'{e}el order at a finite temperature
and they give a good description of the short-distance behavior.
However, we don't claim that these mean-filed states become 
long-range ordered at half filling.
Our interest is in which type of the order
parameter can give the anisotropy in the charge compressibility
at $T \sim J$.
Even if these mean-filed states are unstable,
they can still play an important role at a finite temperature.
 On the other hand,
$\kappa(\bm{k})$ of the N\'{e}el state gives a constant value 
on the FS, which does not reproduce the QMC results.
The ground state of the half-filled system is 
well described by the N\'{e}el state, 
while the system behaves essentially as a non-interacting case
at a sufficiently high temperature ($T \gg U$).
Both the N\'{e}el state and the non-interacting metallic state
give a constant $\kappa(\bm{k})$ on the FS.
Therefore, one of the natural scenarios for the anisotropy in $\kappa (\bm{k})$ is 
that different kinds of fixed points
exist and bring about some singular phenomena
at an intermediate temperature.
Finally, we note that
a possible admixture of different orders may happen
at a low temperature.

 In summary, 
we have investigated the charge fluctuation
in the two-dimensional Hubbard model
by the quantum Monte Carlo method.
The momentum-resolved charge compressibility $\kappa(\bm{k})$
is focused on at a finite temperature.
It gives information on the infinitesimal doping
in the Mott insulator.
The peak structure at the ($\pm\pi/2$,$\pm\pi/2$) points
is observed in $\kappa(\bm{k})$
at the temperature $T \sim {t^2}/{U} $.
It is qualitatively consistent with the calculation for 
the $d$-wave pairing state or the staggered flux state,
while the low temperature($T \ll {t^2}/{U} $) behavior is
dominated by the antiferromagnetic Mott insulator.
This peak structure disappears at the high temperature $T>U$
where the Coulomb interaction is irrelevant.
The crossover observed in our results reflects 
the existence of several fixed points (including unstable ones)
in the strongly correlated electron systems.

\begin{acknowledgments}
We thank
M. Imada, 
Y. Kato,
S. Ryu,
J. Kishine,
K. Yonemitsu,
and P. A. Lee 
for fruitful discussions.
The computation in this work has been done in part
using the facilities of 
the Supercomputer Center, ISSP, University of Tokyo.
\end{acknowledgments}

\end{document}